\documentclass[aps,prb,twocolumn,a4paper,superscriptaddress,floatfix,showpacs,preprintnumbers,amsmath,amssymb]{revtex4-2}
\usepackage{url}
\usepackage[english]{babel}
\usepackage[utf8]{inputenc}
\usepackage{amsmath}
\usepackage{graphicx,epstopdf}
\usepackage{amssymb,epsfig,color,textcase}
\usepackage{hyperref}
\providecommand{\U}[1]{\protect\rule{.1in}{.1in}}

\def\red{\color{red}}

\begin{document}


\title{Ab initio guided minimal model for the ``Kitaev'' material BaCo$_2$(AsO$_4$)$_2$:\\Importance of direct hopping,  third-neighbor exchange and quantum fluctuations}

\author{Pavel A. Maksimov}%
\affiliation{Bogolyubov Laboratory of Theoretical Physics, Joint Institute for Nuclear Research, Dubna, Moscow region 141980, Russia}
\affiliation{M.N. Miheev Institute of Metal Physics of Ural Branch of Russian Academy of Sciences, S. Kovalevskaya St. 18, 620990 Ekaterinburg, Russia}

\author{Alexey V. Ushakov}%
\affiliation{M.N. Miheev Institute of Metal Physics of Ural Branch of Russian Academy of Sciences, S. Kovalevskaya St. 18, 620990 Ekaterinburg, Russia}

\author{Zlata V. Pchelkina}%
\affiliation{M.N. Miheev Institute of Metal Physics of Ural Branch of Russian Academy of Sciences, S. Kovalevskaya St. 18, 620990 Ekaterinburg, Russia}
\affiliation{Department of theoretical physics and applied mathematics, Ural Federal University, Mira St. 19, 620002 Ekaterinburg, Russia}

\author{Ying Li}%
\affiliation{Department of Applied Physics and MOE Key Laboratory for Nonequilibrium Synthesis and Modulation of Condensed Matter, School of Physics, Xi'an Jiaotong University, Xi'an 710049, China}

\author{Stephen M. Winter}%
\affiliation{Department of Physics and Center for Functional Materials, Wake Forest University, NC 27109, USA}

\author{Sergey V. Streltsov}%
\email{streltsov@imp.uran.ru}
\affiliation{M.N. Miheev Institute of Metal Physics of Ural Branch of Russian Academy of Sciences, S. Kovalevskaya St. 18, 620990 Ekaterinburg, Russia}
\affiliation{Department of theoretical physics and applied mathematics, Ural Federal University, Mira St. 19, 620002 Ekaterinburg, Russia}

\date{\today}

\begin{abstract}
By considering two {\it ab initio}-based complementary approaches, we analyze the electronic structure and extract effective spin models of BaCo$_2$(AsO$_4$)$_2$, a honeycomb material which has been proposed as a candidate for Kitaev physics. Both methods show that the dominant direct hopping makes the bond-dependent Kitaev term negligible averting the material away from the sought-after spin-liquid regime. As a result, we present a simple three-parameter exchange model to describe the interactions of the lowest doublet of the honeycomb cobaltate BaCo$_2$(AsO$_4$)$_2$. Remarkably, it is the third-neighbor interactions, both isotropic and anisotropic, that are responsible for the standout double-zigzag ground state of BaCo$_2$(AsO$_4$)$_2$, stabilized by quantum fluctuations. A significantly large third-nearest neighbor hopping, observed in {\it ab initio}, supports the importance of the third-neighbor interactions in the stabilization of the unique ground state of BaCo$_2$(AsO$_4$)$_2$.
\end{abstract}


\maketitle

{\it Introduction.-} In the search for material realizations of Kitaev quantum spin liquids \cite{Kitaev2006}, a whole family of honeycomb cobaltates BaCo$_2$(AsO$_4$)$_2$, BaCo$_2$(PO$_4$)$_2$, CoTiO$_3$ and many others \cite{Khaliullin_Co_2018,Khaliullin_3d_2020,%
Park_Co_2020,Ma_Co_2020,Songvilay_Co_2020,Hess_Co_2021,Wildes_Co_2017,NCTO_zigzag16,NCSO_zigzag16,Zvereva2016,Eymond_1969,Regnault_BaCoAsO_1977,Regnault_1986,Regnault2018,dejongh1990,Regnault_2006,Zhong2020,Wang_THz_2021,Armitage_THz_2021}
were proposed to host dominant nearest-neighbor Kitaev exchange couplings, following the concept that Co$^{2+}$ ions ($3d^7$) in an octahedral crystal field environment with total spin $S=3/2$ and orbital angular momentum $l_\text{eff}=1$ build spin-orbit coupled $j_\text{eff}$= 1/2 doublet states. However, this proposal was recently put into question by a combination
of first-principles-based calculations with single-site exact diagonalization and two-site perturbation theory~\cite{Das2021} where it was found that the Kitaev term
favouring a spin-liquid ground state must be rather small. 

In the present work we perform a detailed analysis of the electronic structure within density functional theory (DFT) employing the projector-augmented wave~\cite{blochl_projector_1994} basis as implemented in VASP, as well as the full potential local orbital (FPLO) basis~\cite{FPLO}. We further use two {\it ab initio}-based complementary approaches to extract effective spin models
for the honeycomb cobaltate BaCo$_2$(AsO$_4$)$_2$. One is based on total energy calculations of various magnetic configurations
 within DFT+U+SOC and mapping into a spin model to extract the exchange tensor elements, and the second method
 is based on the recently introduced projED method~\cite{riedl2019ab,winter2016challenges} by some of the authors
 which consists of a combination of DFT calculations, exact diagonalization (ED) of extracted generalized relativistic Hubbard models on finite clusters, and projection to low-energy spin Hamiltonians.
 We find that the Kitaev model is hardly applicable for a description of the magnetic properties of BaCo$_2$(AsO)$_2$, which is supported by both {\it ab initio} calculations and the phenomenology of BaCo$_2$(AsO)$_2$.
 
One of the peculiarities of BaCo$_2$(AsO$_4$)$_2$ is the long search for the correct magnetic ground state. Early works suggested that the ground state is an unusual long-range ordered spiral state with in-plane ordering vector $\mathbf{Q}=(0,\pi/3)$ \cite{Regnault_BaCoAsO_1977}.  However, more refined neutron scattering data later showed that the ground state is instead a collinear double-zigzag state \cite{Regnault2018} with the $++--$ pattern of zigzag chains with the magnetic moments approximately parallel to the chain direction. Since the model including isotropic exchange between the nearest and 3rd nearest neighbors ($J_1$-$J_3$ model) does not reproduce this experimentally established magnetic order, the Hamiltonian requires extra terms beyond the isotropic third-neighbor interaction in order to describe the ground state of BaCo$_2$(AsO$_4$)$_2$. Indeed, {\it ab initio} studies in this work and in earlier studies \cite{Das2021} indicate a significant role of anisotropic couplings associated with the spin-orbital $j_{\rm eff} = 1/2$ moments of $3d^7$ Co$^{2+}$. This is further supported by the finite energy gap observed in inelastic neutron scattering data and terahertz spectroscopy, although the small magnitude of this gap ($\sim 1.45$ meV \cite{Regnault2018}) places constraints on the magnitude of anisotropic terms.

In this paper we extensively study the phase diagram of the general eight-parameter model to identify which exchange parameters are essential in stabilization of the double-zigzag state. We find that it is the third-neighbor isotropic and anisotropic terms that are crucial and make up a minimal model required to describe the magnetic properties of BaCo$_2$(AsO$_4$)$_2$. Even though the {\it ab initio} parameter sets yield the zigzag ground state, they are close to the phase boundary. Therefore, we argue that there can be additional corrections, such as magnetoelastic coupling, which can tune the calculation towards the experimentally observed double-zigzag ground state.

{\it DFT: crystal-field and hopping parameters.-} We start with the analysis of the on-site Hamiltonian obtained in non-magnetic DFT using both maximally-localized Wannier functions ~\cite{Mostofi2014}
within VASP \cite{kresse_efficient_1996} and FPLO \cite{FPLO,eschrig2004relativistic} basis sets. All calculation details are given in the Supplemental Material (SM) \cite{SM} (see, also, references \cite{kresse_efficiency_1996, kresse_ab_1994,perdew_generalized_1996,Liechtenstein1995,monkhorst_special_1976,Schuler2018,koepernik2021symmetry,Foyevtsova2013,Wien2k,Blaha2020,Dordevic2008,Xiang2011} therein). 

\begin{figure}[t]
\centering
\includegraphics[width=1.0\columnwidth]{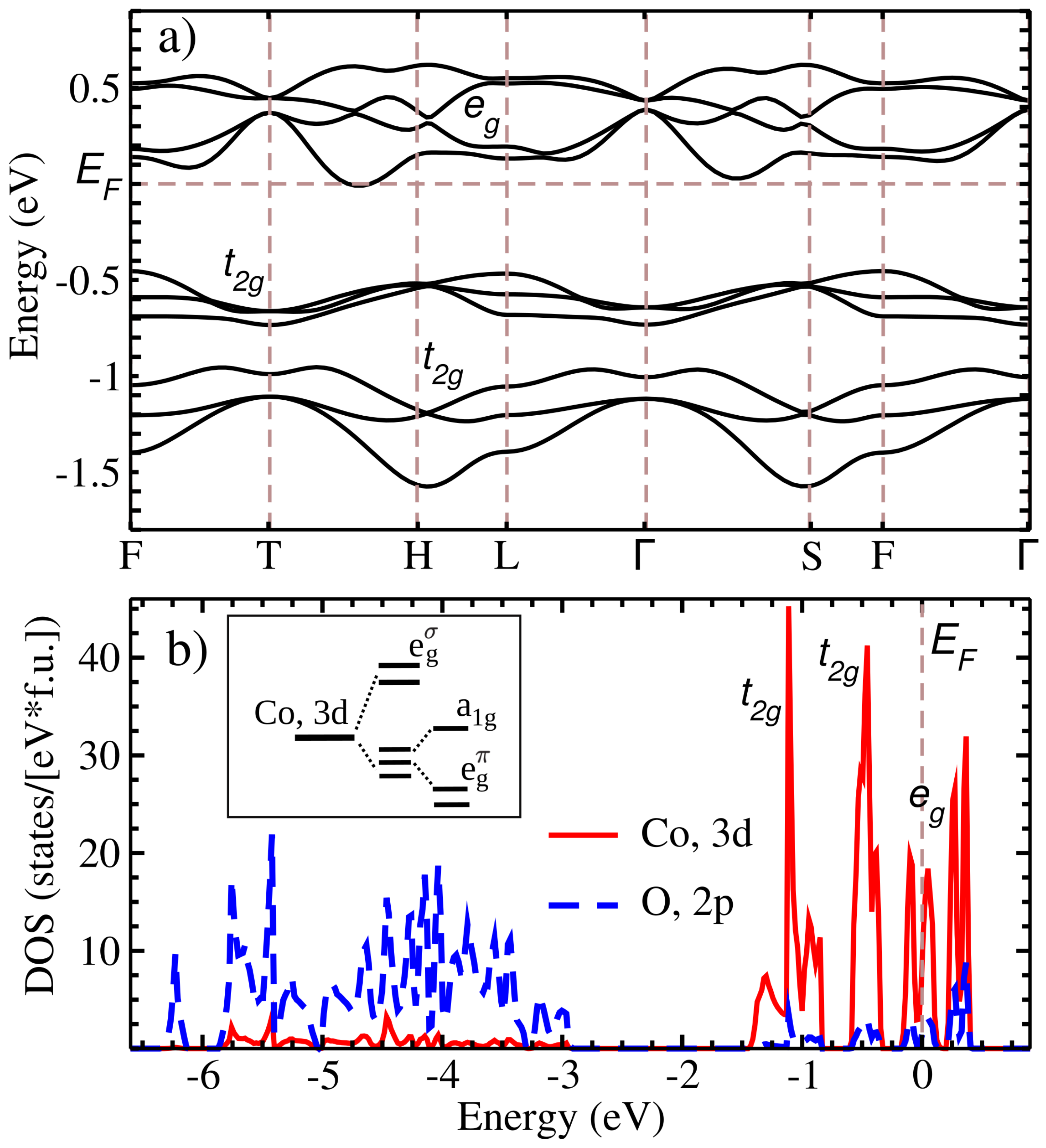}
\caption{(a) The band structure and (b) partial densities of states of BaCo$_2$(AsO$_4$)$_2$ in GGA approximation.}
\label{fig_dosgga}
\end{figure}

\begin{figure}[t!]
\centering
\includegraphics[width=1.0\columnwidth]{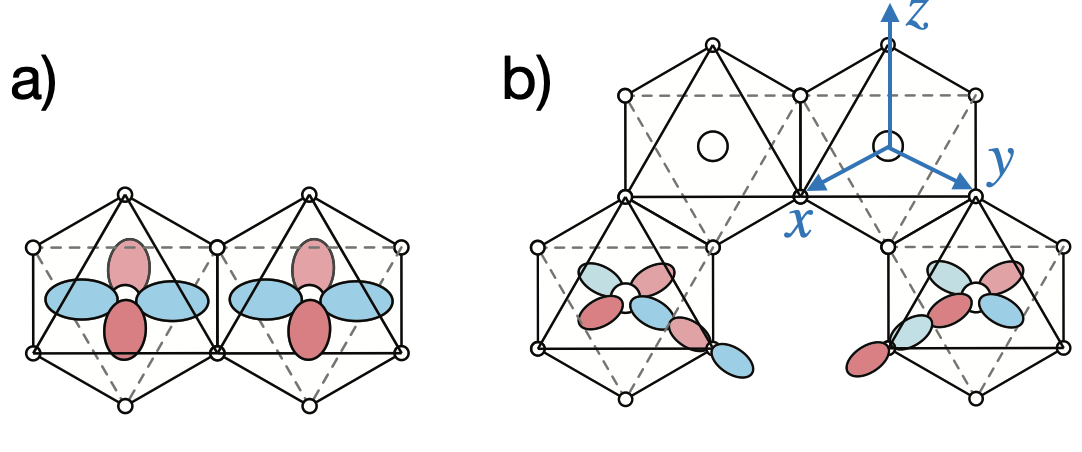}
\caption{Sketch illustrating two dominating hopping processes according to DFT calculations. a) Direct nearest neighbor overlap of two $xy$ orbitals. b) Largest contributions
for third nearest neighbors corresponding to hoppings between $x^2-y^2$ orbitals via O $2p$ states.}
\label{orbitals}
\end{figure}

The crystal structure of BaCo$_2$(AsO$_4$)$_2$ is characterized by the  $R\bar{3}$ space group. It consists of stacked honeycomb layers  of edge sharing CoO$_6$  octahedra along the $c$ axis separated by bilayers of opposite-facing AsO$_4$ tetrahedra with Ba atoms in between. The cubic crystal field of the O$_6$ octahedron around Co splits the Co $3d$ orbitals onto $e_g^{\sigma}$ and $t_{2g}$ states and the latter are then further split by additional trigonal distortions onto lower lying $e_g^{\pi}$ and $a_{1g}$ states because of the layered structure, see inset to Fig.~\ref{fig_dosgga}(b). In Ref.~\onlinecite{Das2021}, these splittings were estimated by rescaling values (to fit available experimental data) obtained from N-th muffin-tin orbital calculations, leading to 735 meV for the $e_{g}^{\sigma} - a_{1g}$ splitting and to 94 meV for the splitting within $t_{2g}$ ($a_{1g}-e_{g}^{\pi}$). Our DFT (VASP) calculations give similar results, 818 meV and 122 meV respectively, without any additional renormalization. Interestingly, with the FPLO basis, Wannier projection~\cite{koepernik2021symmetry} leads to somewhat larger values of 963 meV and 133 meV. However, it should be emphasized that in all cases the latter $t_{2g}$ splitting is comparable to the atomic spin-orbit coupling constant $\lambda_{\rm Co} \approx 60$ meV defined by $\mathcal{H}_{\rm SOC} = \lambda \ \mathbf{S}\cdot \mathbf{L}$, suggesting strong deviations from ideal $j_{\rm eff}={1/2}$ moments at the Co sites. This finding is also consistent with the strong reported anisotropy of the $g$-tensor ($g_{ab} \sim 2 \ g_{c}$) \cite{Regnault2018}. 

We utilized several approaches for the nearest-neighbor hopping integrals: Wannier function projection~\cite{Mostofi2014}, projected localized orbitals (PLO) ~\cite{Schuler2018} as implemented in VASP, and FPLO~\cite{koepernik2021symmetry}. Notably, all approaches we used led to the same conclusions. For instance, for the bond depicted in Fig.~\ref{orbitals}(a), the direct $xy/xy$ hopping $t' \sim -300$ meV dominates over hoppings via $xz$, $yz$ orbitals, $t \sim$ 50 meV, which are associated with electron transfer via ligand $p$ states, local coordinate system shown in Fig.~\ref{orbitals}(b) is chosen. Similar findings were recently reported for Na$_2$BaCo(PO$_4$)$_2$ \cite{wellm2021frustration} and other edge-sharing Co compounds \cite{winter2022tobepublished}. Already this fact reveals that one of the key assumptions used in \cite{Khaliullin_Co_2018}, $\kappa = |t'/t| < 1$, is far from being fulfilled. This results in a small Kitaev exchange for the first-nearest neighbors as we will demonstrate in the next section. It is worth mentioning that a large $xy/xy$ hopping not only affects exchange coupling, but it also strongly changes the electronic structure. One can clearly observe formation of two isolated branches of Co $t_{2g}$ bands in Fig. \ref{fig_dosgga}, where each of $t_{2g}$ orbitals has a direct overlap with the corresponding orbital on one of the neighboring Co sites. Indeed, the gap between these bands closes, if the direct $xy/xy$ hopping is put to zero as explained in SM \cite{SM}.

Another very important feature, which becomes evident already on DFT level, is that hopping between third-nearest neighbors, $t_3$, is not small. The most important contribution comes from hopping between $e_g$ orbitals, which strongly hybridize with ligand $p$ orbitals. This leads, for example, to an effective hopping $t_3^{x^2-y^2/x^2-y^2} \approx 124$  meV, see Fig.~\ref{orbitals}(b). 
This hopping is associated with an antiferromagnetic $\text{J}_3$ that may be comparable to the nearest neighbor exchange $\text{J}_1$. This finding is in contradiction with the assumption \cite{Khaliullin_Co_2018,Motome2020,Khaliullin_3d_2020} that longer range couplings should be suppressed in $3d^7$ compounds as a result of stronger Coulomb interactions and smaller $d$ orbitals localizing the moments in comparison to traditional $4d^5$ and $5d^5$ Kitaev candidate materials. In fact, the partial filling of the $e_g$ orbitals in $3d^7$ Co$^{2+}$ provides additional long-range exchange pathways.  Large hopping between the third-nearest neighbors is also reflected in electronic structure of BaCo$_2$(AsO$_4$)$_2$. It results in formation of the bonding and antibonding $e_g$ states, as discussed in SM \cite{SM}, which are clearly seen in Fig.~\ref{fig_dosgga}. 
\begin{table}[t]
\centering
\caption{The dependence of exchange interaction parameters on on-site Coulomb $U$ computed from DFT+SOC+U total energy in the extended Kitaev model. Intra-atomic Hund's exchange was chosen to be $J_H$=0.9~eV.}
\begin{tabular}{l | c c c}
\hline
\hline
 $U$ & 5 eV  & 6 eV & 7 eV \\
\hline 
 $\text{J}_1$ (K)                & -61.0   & -40.9   &  -37.6  \\
 $K_1$ (K)              &  0.3   & 2.2  & 5.3  \\ 
$\Gamma_1$  (K)          &  -2.2  & -1.7  &  -1.8 \\ 
  $\Gamma_{1}^{\prime}$ (K) & 5.1  & 4.0  & 3.2 \\ 
\hline
 $\text{J}_3$ (K)             & 31.4  &  24.6  &  18.7  \\
   $K_3$ (K)           &  -0.2 & 0.2  & -0.2  \\
$\Gamma_3$   (K)       &  -4.5  & -6.0  & -4.5  \\
$\Gamma^{\prime}_{3}$ (K) & -3.6   & -2.3  & -1.8   \\
\hline
\end{tabular}
\label{JPavel}
\end{table}
\begin{table}[t]
\centering
\caption{The dependence of exchange interaction parameters on on-site Coulomb $U$ in the crystallographic parameterization. Intra-atomic Hund's exchange was chosen to be $J_H$=0.9~eV.}
\begin{tabular}{l | c c c}
\hline
\hline
 $U$ & 5 eV  & 6 eV & 7 eV \\
\hline 
  $J_1$   (K)              &  -63.6   & -42.3  & -37.4  \\ 
$\Delta_1$            &  0.87 & 0.85 &  0.88\\ 
  $J_{\pm\pm}^{(1)}$  (K)      & 2.4  & 1.5  & 0.8 \\ 
 $J_{z\pm}^{(1)}$  (K)         & 3.5   & 3.7   &  4.9  \\
\hline
  $J_3$  (K)               &  35.2   & 28.2  & 21.3  \\ 
$\Delta_3$            &  0.67 & 0.62 &  0.62\\ 
  $J_{\pm\pm}^{(3)}$ (K)       & 0.3  & 1.2  & 0.9 \\ 
 $J_{z\pm}^{(3)}$  (K)         & 0.3   & 1.8   &  1.2  \\
\hline
\hline
\end{tabular}
\label{J}
\end{table}

\begin{figure}
\centering
\includegraphics[width=1.0\columnwidth]{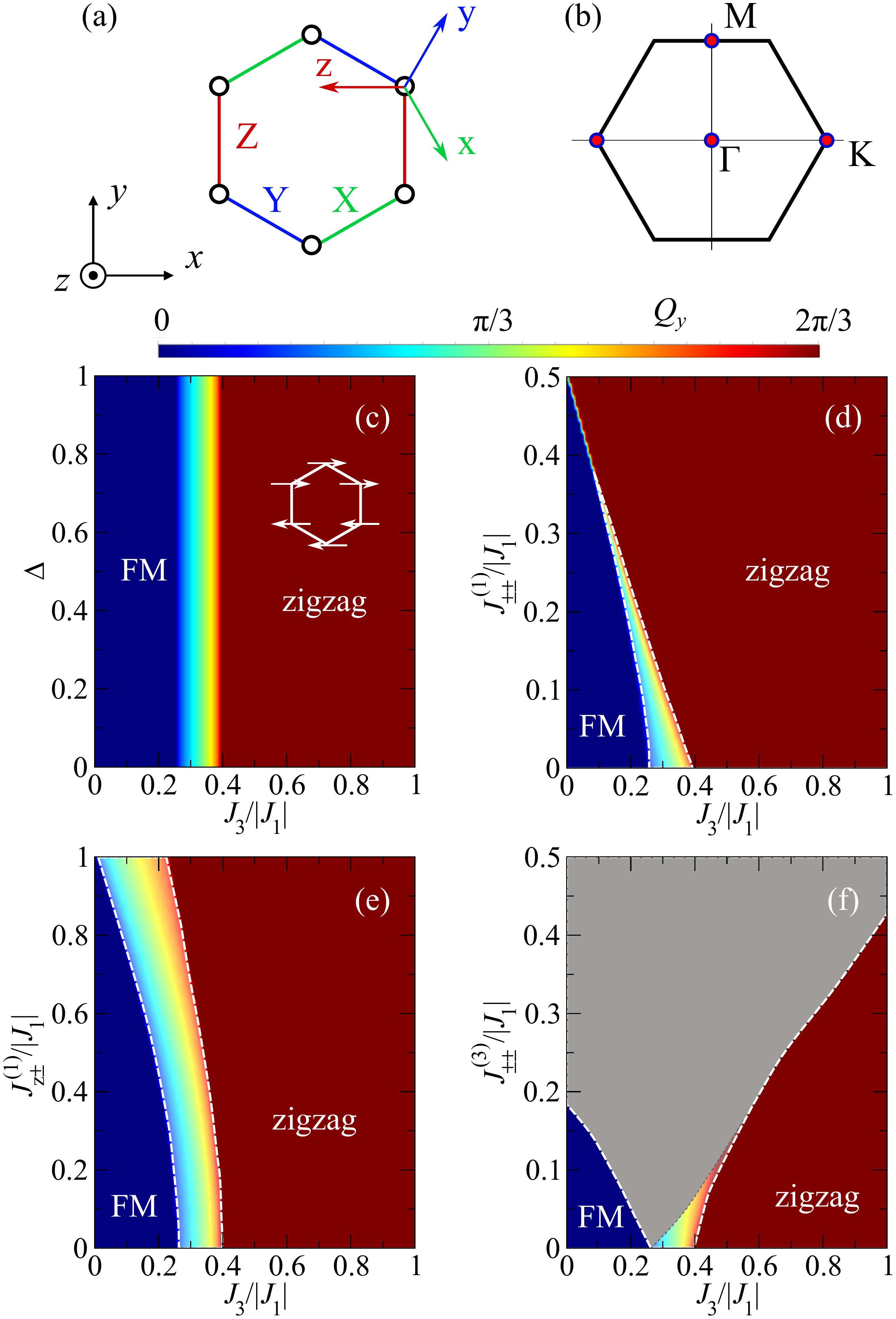}
\caption{(a) Honeycomb lattice with three types of bonds; cubic axes \{x,y,z\} and crystallographic axes $\{x,y,z\}$ are also presented. (b) Brillouin zone of the honeycomb lattice with high-symmetry points indicated. (c)-(f) Classical phase diagram of the model \eqref{HJpm} calculated with Luttinger-Tisza method for $J_1<0$, $J_3>0$, $\Delta = \Delta_1 = \Delta_3$. The ordering vector of the classical spiral state $Q_y$ is represented with the color intensity map. A sketch of the zigzag shown as an inset. The Spiral-II state has non-zero $Q_x$ component of the ordering vector, thus not related to the ground state of BaCo$_2$(AsO$_4$)$_2$.}
\label{fig_lt}

\end{figure}

{\it Ab initio Exchange Parameters.-} 
Due to the interplay of crystal field and SOC, the magnetic interactions of effective doublets are defined by the symmetry of the lattice. The symmetry of edge-sharing CoO$_6$ octahedra  results in the extended Kitaev-Heisenberg model \cite{rau2014trigonal}, which  has been discussed for BaCo$_2$(AsO$_4$)$_2$ \cite{Khaliullin_3d_2020}. Explicitly the exchange Hamiltonian is given by
\begin{align}
\hat{\cal H}_\text{cubic}=\sum_{ \langle ij \rangle_n }&
\text{J}_n \ \mathbf{S}_i \cdot \mathbf{S}_j +K_n S^\gamma_i S^\gamma_j 
+\Gamma_n \left( S^\alpha_i S^\beta_j +S^\beta_i S^\alpha_j\right)\nonumber\\
+&\Gamma'_n \left( S^\gamma_i S^\alpha_j+S^\gamma_i S^\beta_j+S^\alpha_i S^\gamma_j+S^\beta_i S^\gamma_j\right),
\label{eq_H_JKGGp}
\end{align}
where the sum is taken over the three types of bonds of the honeycomb lattice, $\{\alpha,\beta,\gamma\}=\{\text{x,y,z}\}$ for the Z-type bond and interactions on the X and Y bonds are obtained through a cyclic permutation \cite{rau_jkg,rau2014trigonal}, see Fig.~\ref{fig_lt}(a). Note that this model uses cubic axes \{x,y,z\}, which are shown in Figs.~\ref{orbitals} and \ref{fig_lt}(a) and related to the ion-ligand bonds. In accordance with neutron studies, and {\it ab initio} results below, we anticipate that first-neighbor ($n = 1$) and third-neighbor ($n=3$) couplings are dominant. 

For our purpose, it is also convenient to refer the interactions alternatively to the crystallographic axes $\{x,y,z\}$ which are defined by the honeycomb plane of magnetic ions, shown in Fig.~\ref{fig_lt}(a). The Hamiltonian in that reference frame is given by
\begin{align}
\hat{\cal H}_\text{cryst}=&\sum_{\langle ij \rangle_n}
J_n \Big(S^{x}_i S^{x}_j+S^{y}_i S^{y}_j+\Delta_n S^{z}_i S^{z}_j\Big)\nonumber\\-&2 J_{\pm \pm}^{(n)} \Big( \Big( S^x_i S^x_j - S^y_i S^y_j \Big) c_\alpha 
-\Big( S^x_i S^y_j+S^y_i S^x_j\Big)s_\alpha \Big)\nonumber\\
 -&J_{z\pm}^{(n)}\Big( \Big( S^x_i S^z_j +S^z_i S^x_j \Big) c_\alpha 
 +\Big( S^y_i S^z_j+S^z_i S^y_j\Big)s_\alpha \Big),
\label{HJpm}
\end{align}
where $c_\alpha\equiv\cos\varphi_\alpha$ and $s_\alpha\equiv\sin\varphi_\alpha$ with the bond-dependent phases $\varphi_\alpha\!=\!\{0,2\pi/3,-2\pi/3\}$ for the three types of first and third-neighbor bonds \cite{rau2014trigonal,chaloupka2015}. The exchange parameters of the extended Kitaev-Heisenberg model \eqref{eq_H_JKGGp} and anisotropic-exchange model in the crystallographic axes \eqref{HJpm} are related through a simple linear transformation:
\begin{align}
\text{J}_n&=\frac{1}{3}\left( 2J_n+\Delta_n J_n+2J_{\pm\pm}^{(n)}-\sqrt{2} J_{z\pm}^{(n)}\right),\nonumber\\
\label{eq_jkg_transform}
K_n&=-2J_{\pm\pm}^{(n)}+\sqrt{2}J_{z\pm}^{(n)},\\
\Gamma_n&=\frac{1}{3} \left( -J_n+\Delta_n J_n-4J_{\pm\pm}^{(n)}-\sqrt{2} J_{z\pm}^{(n)}\right),\nonumber\\
\Gamma'_n&=\frac{1}{6} \left( -2J_n+2\Delta_n J_n+4J_{\pm\pm}^{(n)}+\sqrt{2} J_{z\pm}^{(n)}\right).\nonumber
\end{align}

We use a variety of approaches to estimate the magnetic exchange couplings.  Conventional DFT calculations underestimate the effect of strong Coulomb correlations, which must be taken into account for extraction of the exchange interaction in the transition metal oxides. We therefore first computed the magnetic interactions based on DFT(GGA)+U+SOC calculation of total energies of four non-colinear magnetic configurations~\cite{Xiang2011}.  These results are complemented by extraction of the couplings from exact diagonalization of two-site clusters, described below. 

For the total energy DFT(GGA)+U+SOC calculations, the most important exchanges for the first and third-nearest neighbors are summarized in Tables~\ref{JPavel} and \ref{J} in terms of both extended Kitaev and crystallographic parameterizations (other constants were found to be small, e.g. $J_2 \sim 0.2$K). The calculations were performed for several values of the on-site Hubbard repulsion parameter $U$, but all of them demonstrate that (i) the Kitaev exchange is small both for the first and third-nearest neighbors and that (ii) there is a strong exchange coupling with third-nearest neighbors. Both factors strongly suppress formation of a spin-liquid state, and are compatible with previous neutron scattering analysis, which suggested $J_1 \sim -38$~K, $J_3 \sim +10$ K \cite{Regnault_1986}, as well as more recent data \cite{Broholm_BCAO_2022}, which estimated $J_1 \sim -88$ K, and $J_3 \sim +29$ K, and $J_{\pm\pm}^{(1)} \sim -0.6$ K. We note, however, that large XXZ anisotropy estimated from experiment ($\Delta \sim 0.37$ \cite{Regnault_1986} and $\Delta_1 \sim 0.16$ \cite{Broholm_BCAO_2022}), is not reproduced in this approach.


In order to further examine the magnetic couplings, we employed a complementary approach similar to Ref.~\onlinecite{Das2021}: exact diagonalization of the five $d$-orbital model on two sites~\cite{riedl2019ab,winter2016challenges}. For this purpose, we employ hopping integrals obtained from VASP, and take the fully spherically symmetric form~\cite{sugano2012multiplets} of the on-site Coulomb interactions with Slater parameters $F_4/F_2 = 0.625$, and $F_0 \equiv U$ and $F_2 \equiv 14J/(1+0.625)$ set according to $U = 5$ to 7 eV, and $J_{H,t_{2g}} = 0.9$ eV. The results are shown in Table~\ref{JED}. For the purpose of comparison, results for Coulomb parameters equivalent to Ref.~\onlinecite{Das2021} ($U = 3.25$ eV, $J_{H,t_{2g}} = 0.7$ eV) are also shown. We note that this approach neglects an important contribution to the exchange involving multiple holes on a given ligand, which can be corrected using expressions from perturbation theory \cite{winter2022tobepublished,lines1963magnetic,Khaliullin_3d_2020}. This leads to shifts of the nearest neighbor couplings $\text{J}_1 \to \text{J}_1+\delta \text{J}$, and $\Gamma_1 \to \Gamma_1 + \delta \Gamma$ and $\Gamma_1^\prime \to \Gamma_1^\prime + \delta \Gamma$ that can be estimated from the trigonal crystal field splitting and realistic metal-ligand hopping parameters (see \cite{winter2022tobepublished, Khaliullin_3d_2020}). In this case, a rough estimate is $\delta J \sim -20$ K, and $\delta \Gamma \sim +7$ K. Both corrected and uncorrected results are given in Tables \ref{JED} and \ref{pavelED}.
\begin{table}[t!]
\centering
\caption{Exchange interaction parameters (in units of K) computed from exact diagonalization of effective $d$-orbital model. Values in brackets include estimated corrections for omitted ligand exchange processes.}
\begin{tabular}{l | c | c c c}
\hline
\hline
$J_{H,t_{2g}}$&0.7~eV&\multicolumn{3}{c}{0.9~eV}\\
 $U$&3.25 eV & 5 eV  & 6 eV & 7 eV \\
\hline 
 $\text{J}_1$ (K)                & -107 (-127)& -37 (-57)  & -18 (-38)  &  -8.8 (-29) \\
  $K_1$ (K)               & 32 &  13   &6.5 & 3.4 \\ 
$\Gamma_1$ (K)            & 28 (35)&  14 (21) & 8.0 (15) &  4.8 (12)\\ 
  $\Gamma_1^{\prime}$ (K) & 9.4 (16) & 7 (14) & 4.0 (11) & 2.4 (9) \\ 
\hline
$\text{J}_3$ (K)       & 43 & 30   &27  & 24
\\
$K_3$ (K)              &-0.6 & -0.4 &-0.3 & -0.3
\\
$\Gamma_3$ (K)         &-20 & -12  & -10& -8.9
\\
$\Gamma_3^\prime$ (K)  & -21 & -12  & -11 & -9.2
\\
\hline
\hline
\end{tabular}
\label{JED}
\end{table}

\begin{table}[t!]
\centering
\caption{Exchange interaction parameters computed from exact diagonalization of effective $d$-orbital model. Values in brackets include estimated corrections for omitted ligand exchange processes.}
\begin{tabular}{l | c | c c c}
\hline
\hline
$J_{H,t_{2g}}$&0.7~eV&\multicolumn{3}{c}{0.9~eV}\\
 $U$&3.25 eV & 5 eV  & 6 eV & 7 eV \\
\hline 
 $J_1$ (K)               & -113 (-140)& -42 (-69)  & -21 (-48)  &  -11 (-38) \\
  $\Delta_1$        & 0.58 (0.51) &  0.36 (0.31)   &0.24 (0.23) & 0.11 (0.19) \\ 
$J_{\pm\pm}^{(1)}$ (K)     & -12&  -4.5 & -2.4&  -1.4\\ 
  $J_{z\pm}^{(1)}$ (K)   & 6.2 & 2.7  & 1.2 & 0.4 \\ 
\hline 
 $J_3$ (K)               & 64& 42  & 37  &  33 \\
  $\Delta_3$        & $\sim 0$ &  0.14   &0.16 & 0.18 \\ 
$J_{\pm\pm}^{(3)}$ (K)     & 0.1&  $\sim 0$ & $\sim 0$ &  $\sim 0$\\ 
  $J_{z\pm}^{(3)}$ (K)   & -0.60 & -0.36  & -0.30 & -0.26 \\
\hline
\hline
\end{tabular}
\label{pavelED}
\end{table}

The ED + perturbation theory results are essentially compatible with the DFT results above; ferromagnetic $J_1$ dominates the couplings, while the anisotropic couplings $K_1,\Gamma_1, \Gamma_1^\prime > 0$ are all of smaller but similar magnitude. The main difference is that the XXZ anisotropy is considerably stronger in ED results, with corrected values of $\Delta_1$ ranging between $0.2$ and $0.5$. This anisotropy originates from the effects of local trigonal crystal field on the $j_{\rm eff} = 1/2$ multiplet structure, which may not be completely captured in one-electron methods such as DFT+U+SOC. On the other hand, cluster approaches, such as our ED method, tend to have difficulties with long-range interactions. This fact is embodied by surprisingly large $J_3$, which is comparable to $J_1$, as can be seen in Table~\ref{pavelED}.

Thus, both calculation methods rather guide than provide exact estimates of exchange constants. It is important that our methods and perturbation theory of Ref.~\cite{Das2021} give consistent results, which is not always the case for Kitaev materials (for Li$_2$IrO$_3$ see Ref.~\cite{Nishimoto2016,Winter2016} and for $\alpha$-RuCl$_3$ see Ref.~\cite{us_PRR}). However, both approaches place the system in a region of zigzag order for all of the presented parameter sets, rather than the double-zigzag observed in BaCo$_2$(AsO$_4$)$_2$. In the next section we show that these parameter sets are, in fact, close to the boundary of the double-zigzag phase, and extensively study the full phase diagram in order to establish which of the parameters of the model would tune the ground state towards the experimentally established one.

\begin{figure}
\centering
\includegraphics[width=1.0\columnwidth]{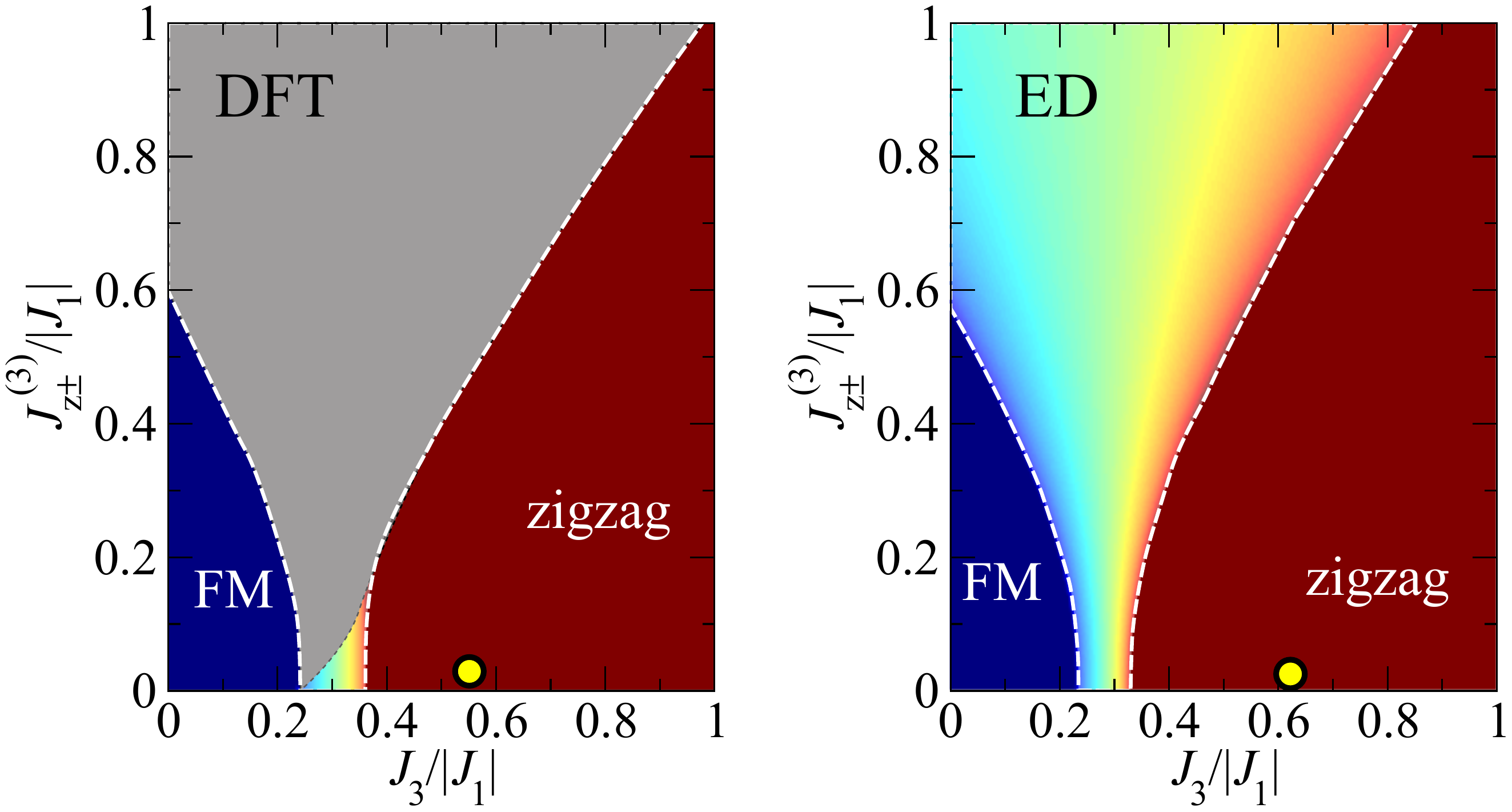}
\caption{Luttinger-Tisza $J_3$-$J_{z\pm}^{(3)}$ phase diagrams for $U=5$ eV sets of parameters from Tables \ref{JPavel} and \ref{JED}. DFT and ED parameter sets are indicated with yellow dots.}
\label{fig_ed_dft_pd}
\end{figure}

\begin{figure*}
\centering
\includegraphics[width=2.0\columnwidth]{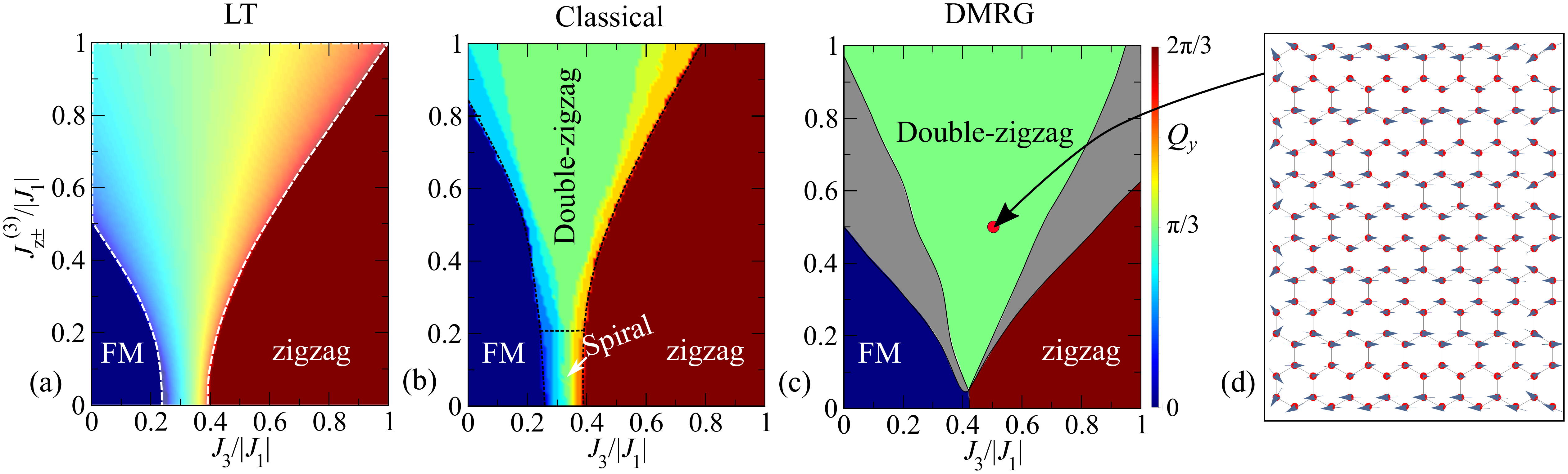}
\caption{(a)  Luttinger-Tisza phase diagram of the $XY$ (i.e.~$\Delta = 0$) $J_1$-$J_3$-$J_{z\pm}^{(3)}$ model \eqref{eq_ham} calculated with Luttinger-Tisza method for $J_1<0$, $J_3>0$. The ordering vector of the classical state $Q_y$ is represented with the color intensity map. (b) Classical phase diagram of the model \eqref{eq_ham} with the same notation for the intensity map. (c) DMRG phase diagram of the model \eqref{eq_ham} for $S=1/2$. The gray areas indicate regions of the intermediate phases which are beyond the scope of this work. (d) Example of DMRG calculation for a representative parameter set $J_3=J_{z\pm}^{(3)}=0.5|J|$ which exhibits a double-zigzag ground state.}
\label{fig_phasediagram}
\end{figure*}

{\it Phase Diagram.-} We now investigate the minimal set of interactions that is compatible with the ground state phenomenology of BaCo$_2$(AsO$_4$)$_2$ \cite{dejongh1990}. We start with the isotropic $XY$ $J_1$-$J_3$ model, guided by {\it ab initio} sets presented above, and study phase transitions when the anisotropic-exchange terms of the full eight-parameter model \eqref{HJpm} are added to the Hamiltonian one at a time, using the  classical Luttinger-Tisza (LT) method \cite{lt_original}. We should note that in frustrated magnets with broken continuous symmetry, incommensurate spiral states given by Luttinger-Tisza approach cannot satisfy local spin constraint, thus are not exact classical ground states \cite{Chen16}. However, even though LT method breaks down, it can still point to regions of exotic ground states on the phase diagram \cite{Ioannis_JK,rau_jkg,us_anisotropic}. Therefore, we use LT method in this section as guide for the search of classical ground states beyond ferromagnetic and zigzag phases by plotting the ordering vector selected by LT, and we take the boundary of stability of LT method as an indicator of the regions of the phase diagram which can host the double-zigzag state (which is studied in detail in the next section).

Corresponding phase diagrams are shown in Fig.~\ref{fig_lt} where we plot the ordering vector $(0,Q_y)$ as an intensity map, and the regions where LT method breaks down are shown with white dashed lines. As depicted in Brillouin zone in Fig.~\ref{fig_lt}(b), $Q_y=0$ corresponds to a ferromagnetic (FM) state, $Q_y=2\pi/3$ to a zigzag state, and  the spiral state originally proposed for BaCo$_2$(AsO$_4$)$_2$ interpolates between these states with $0<Q_y<2\pi/3$. The $J_1$-$J_3$ model is known to host this spiral state for $0.25|J_1|<J_3<0.4|J_1|$ \cite{Rastelli_1979}, as shown in Fig.~\ref{fig_lt}(c), and the addition of XXZ anisotropy  (with $\Delta$ = $\Delta_1$=$\Delta_3$) does not alter the relative stabilities of the phases compared to the pure Heisenberg $J_1$-$J_3$ model. 

In contrast, the addition of anisotropic first-neighbor $J_{\pm\pm}^{(1)}$ and $J_{z\pm}^{(1)}$ exchange terms stabilizes zigzag state, see Fig.~\ref{fig_lt}(d-e). Whereas, the $J_{\pm\pm}^{(3)}$ interaction promotes non-FM and non-zigzag states, shown in gray in Fig.~\ref{fig_lt}(f). However, LT suggests states with the ordering vector near the K point with non-zero $Q_x$ component, thus this region is not relevant to the ground state of BaCo$_2$(AsO$_4$)$_2$ and we do not go into the specifics of its structure.

Finally, we also considered phase diagrams starting from the DFT and ED parameters. For this purpose, we use the results for $U = 5$ eV in each case, as the corresponding interactions are closest in magnitude to the experimental estimates \cite{Regnault_1986,Broholm_BCAO_2022}. Results are shown in Fig.~\ref{fig_ed_dft_pd} with respect to tuning $J_3$ and $J_{z\pm}^{(3)}$. The latter 3rd neighbor interaction was selected  because it tends to stabilize the $(0,Q_y)$ spiral, when added to the both the isotropic $J_1-J_3$ and ED models. As can be seen, both {\it ab initio} parameter sets predict zigzag as a ground state, but both lie relatively close to the boundary of zigzag state. While we find that various modifications of these parameters may stabilize the spiral phase (such as increasing $J_{z\pm}^{(3)}$ with respect to the ED couplings), the discrepancy is most likely attributable to an overestimation of third neighbor couplings. That is, rescaling of $J_3$ shifts both parameter sets towards the spiral state. As we show in the next section, the spiral is indeed replaced by the unique double-zigzag state in a more careful study of the classical ground state and when quantum fluctuations are taken into consideration.


{\it Double-zigzag state.-}  In order to search for the double-zigzag state and explore quantum effects on the relevant phases, we focus on a reduced three-parameter $J_1$-$J_3$-$J_{z\pm}^{(3)}$ model as a minimal Hamiltonian for BaCo$_2$(AsO$_4$)$_2$:
\begin{align}
\hat{\cal H}_\text{min}=&\sum_{\langle ij\rangle_1}
J_1 \Big(S^{x}_i S^{x}_j+S^{y}_i S^{y}_j\Big) +J_3 \sum_{\langle ij\rangle_3}\Big(S^{x}_i S^{x}_j+S^{y}_i S^{y}_j\Big) \nonumber\\
 -&J_{z\pm}^{(3)}\Big( \Big( S^x_i S^z_j +S^z_i S^x_j \Big) c_\alpha 
 +\Big( S^y_i S^z_j+S^z_i S^y_j\Big)s_\alpha \Big),
\label{eq_ham}
\end{align}
We take easy-plane anisotropy $\Delta_1=\Delta_3=0$. This choice of minimal interactions is sufficient to provide a wide region of {\red stability for the double-zigzag} phase, although it neglects many interactions relevant to the real material. 


First, the LT phase diagram of the model \eqref{eq_ham} presented in Fig.~\ref{fig_phasediagram}(a), points to a state with the $(0,Q_y)$ ordering vector. However, since the incommensurate spiral state given by LT does not satisfy the strong local spin length constraint, we perform a more focused study of the classical ground state. We use a quasi-one-dimensional cluster, periodic in $x$-axis, using the fact that spin modulation is only along the $y$-axis, as suggested by the $(0,Q_y)$-spiral state from LT calculations. The results of classical energy minimization on a 24-unit cell cluster are shown in Fig.~\ref{fig_phasediagram}(b), where color intensity illustrates the magnitude of $Q_y$ ordering vector. One can see that, compared to LT method, double-zigzag state is in fact stabilized in the phase diagram of the $J_1$-$J_3$-$J_{z\pm}^{(3)}$ model for $J_{z\pm}^{(3)} \gtrsim 0.2$, while the spiral state is stable for $J_{z\pm}^{(3)} \lesssim 0.2$.

Moreover, it is known that there are strong renormalizations of phase diagrams of frustrated quantum $S=1/2$ models relative to classical $S\rightarrow \infty$ approximation \cite{topography,rau_jkg,zhu_white15,Vojta2020_NLSWT,iqbal16_j1j2,Fouet2001,Fisher_2013,Fisher_2014,Sheng_kagome,Chaloupka_jkg_2019,Suzuki2021}. In order to study the model \eqref{eq_ham} in the quantum limit in the context of $j_\text{eff}=1/2$ moments of BaCo$_2$(AsO$_4$)$_2$, we employ density matrix renormalization group (DMRG) \cite{white_density_1992} using ITensor library \cite{itensor} on a 192-site $S=1/2$ cluster with open boundary conditions using 20 sweeps with error $<10^{-4}$ and a random initial state. (We have also studied clusters of other sizes and with periodic boundary conditions, they all yielded very similar results.) The phases were identified by the maximum value of spin-spin correlator $\mathcal{S}(\mathbf{k})$ calculated at $\Gamma$, $K$, $M$, $(0,\pi/3)$ where
\begin{align}
\mathcal{S}(\mathbf{k})=\sum_{i,j} \langle \mathbf{S}_i\cdot \mathbf{S}_j\rangle e^{i\mathbf{k}\left( \mathbf{r}_i-\mathbf{r}_j\right)}.
\end{align}

The phase diagram, obtained with DMRG in the quantum $S=1/2$ limit, is shown in Fig.~\ref{fig_phasediagram}(c). The ordering vector is shown with the color intensity, same as in Figs.~\ref{fig_phasediagram}(a) and \ref{fig_phasediagram}(b), the phases in gray are intermediate between FM, double-zigzag and zigzag but their characterization is beyond the scope of this work. Note that our investigative DMRG calculation is unable to give a conclusive result in the region of multiple phase competition for $J_{z\pm}^{(3)}<0.1|J_1|$.  Nonetheless, one can see that the double-zigzag state is stable in a wider region of the phase diagram for $S=1/2$, relative to classical model predictions. This fact implies that quantum fluctuations play significant role, which is captured in DMRG. This mechanism is generic, and applies beyond the minimal model. Such fluctuations are known to stabilize collinear orders in frustrated systems, such as the field-induced up-up-down state in the triangular lattice antiferromagnet \cite{chubukov91,yamamoto}, honeycomb $J_1$-$J_2$ model \cite{bishop4}, and anisotropic-exchange model on a triangular lattice \cite{multiQ,topography,us_anisotropic}.

An example of the spin orientations obtained by DMRG for the representative parameter set $J_3=J_{z\pm}^{(3)}=0.5|J_1|$ is shown in Fig.~\ref{fig_phasediagram}(c). This observed spin structure is precisely the same as measured in the latest neutron data \cite{Regnault2018}; the $++--$ double-zigzag structure. Moreover, an out-of-plane canting of the spins around $5^\circ$ was also reported \cite{Regnault2018}. We also observe the out-of-plane canting, induced by the anisotropic $J_{z\pm}^{(3)}$ term, which couples in-plane and out-of-plane spin components. However, the canting in our DMRG calculation is not only of opposite sign between the chains of opposite direction but also has different sign between the A and B sublattices of the honeycomb lattice.

Finally, we remark on an additional mechanism that may stabilize the mysterious double-zigzag structure. Our DFT+U+SOC calculations ($U=6$ eV) show that zigzag is the ground state magnetic structure with the double-zigzag being 1.2 meV/f.u. higher in energy for the experimental crystal structure. However, relaxation of the atomic positions completely changes the situation: the double-zigzag order becomes more stable than the spiral by 0.2 meV/f.u. ($\approx 1$K/Co). The task of deciphering the origin of the stabilization of the double-zigzag structure from \textit{ab initio} is extremely complicated due to a tiny total energy difference (corresponding hopping parameters and nearest neighbor exchange constants can be found in SM \cite{SM}), but one can conclude with certainty that (i) 
the system is on the border between two magnetic phases and phase separation or presence of different domains are not excluded; (ii) the magneto-elastic coupling is important in BaCo$_2$(AsO$_4$)$_2$ \cite{Budko_BCAO}.

{\it Conclusions.-} By means of {\it ab initio} band structure, Luttinger-Tisza and DMRG calculations we studied the electronic and magnetic properties of BaCo$_2$(AsO$_4$)$_2$ - a candidate material for the realization of the celebrated Kitaev model. While previous theoretical results \cite{Khaliullin_3d_2020} and experimental data \cite{Armitage_THz_2021} suggested dominant Kitaev interaction, promising proximity to the spin-liquid regime, in this paper we showed that this notion is not supported neither by phenomenology, nor by various {\it ab initio} methods. 

Instead, DFT and ED in this paper establish that large direct exchange due to $t_{2g}$ orbitals strongly suppresses anisotropic contributions to the exchange interaction between nearest neighbors in agreement with Ref.~\onlinecite{Das2021}. Moreover, there is also substantial coupling with third-nearest neighbors, such that $\text{J}_3 > 0.3 | \text{J}_1|$. These two findings together make the  formation a spin-liquid state unfavorable, driving the system towards the long-range ordered state.  

However, proposed {\it ab initio} models do not yield the unique double-zigzag ground state of BaCo$_2$(AsO$_4$)$_2$. Through the extensive search over the eight-parameter phase space we establish a minimal model that hosts double-zigzag state in a wide range of parameters. Remarkably, quantum fluctuations inherent to $j_\text{eff}=1/2$ magnets strongly affect the ground state of the proposed model and stabilize the double-zigzag magnetic structure previously observed experimentally.  We show that  proposed parameter sets from {\it ab initio} are near the boundary to double-zigzag state and we suggest that magnetoelastic coupling can play a crucial role since double-zigzag state can be stabilized by optimization of the crystal structure in DFT+U+SOC calculations.


{\it Acknowledgments.} We cannot describe the immeasurable gratitude towards Roser Valent\'i, whose expertise immensely assisted in the launch of this project, and whose numerous insightful discussions aided in guiding the paper in the desirable direction. We would like to thank Sasha Chernyshev and Arun Paramekanti for useful discussions and comments. S.S. is grateful to B. Cava and I. Solovyev for fruitful communications. We acknowledge support of the Russian Science Foundation via project  20-62-46047.



\bibliography{calc-details}

\end{document}



\title{Ab initio guided minimal model for the ``Kitaev'' material BaCo$_2$(AsO$_4$)$_2$:\\Importance of direct hopping,  third-neighbor exchange and quantum fluctuations\\Supplemental Materials}

\author{Pavel A. Maksimov}%
\affiliation{Bogolyubov Laboratory of Theoretical Physics, Joint Institute for Nuclear Research, Dubna, Moscow region 141980, Russia}
\affiliation{M.N. Miheev Institute of Metal Physics of Ural Branch of Russian Academy of Sciences, S. Kovalevskaya St. 18, 620990 Ekaterinburg, Russia}

\author{Alexey V. Ushakov}%
\affiliation{M.N. Miheev Institute of Metal Physics of Ural Branch of Russian Academy of Sciences, S. Kovalevskaya St. 18, 620990 Ekaterinburg, Russia}

\author{Zlata V. Pchelkina}%
\affiliation{M.N. Miheev Institute of Metal Physics of Ural Branch of Russian Academy of Sciences, S. Kovalevskaya St. 18, 620990 Ekaterinburg, Russia}
\affiliation{Department of theoretical physics and applied mathematics, Ural Federal University, Mira St. 19, 620002 Ekaterinburg, Russia}

\author{Ying Li}%
\affiliation{Department of Applied Physics and MOE Key Laboratory for Nonequilibrium Synthesis and Modulation of Condensed Matter, School of Physics, Xi'an Jiaotong University, Xi'an 710049, China}

\author{Stephen M. Winter}%
\affiliation{Department of Physics and Center for Functional Materials, Wake Forest University, NC 27109, USA}

\author{Sergey V. Streltsov}%
\email{streltsov@imp.uran.ru}
\affiliation{M.N. Miheev Institute of Metal Physics of Ural Branch of Russian Academy of Sciences, S. Kovalevskaya St. 18, 620990 Ekaterinburg, Russia}
\affiliation{Department of theoretical physics and applied mathematics, Ural Federal University, Mira St. 19, 620002 Ekaterinburg, Russia}

\date{\today}

\date{\today}

\begin{abstract}
 
\end{abstract}

\pacs {71.27.+a, 71.20.-b, 71.15.Mb}

\maketitle

\section{DFT calculation details} 

We carried out an {\it ab initio} band structure calculations within the frameworks of density functional theory (DFT) as implemented in VASP~\cite{kresse_efficient_1996, kresse_efficiency_1996, kresse_ab_1994} and a full-potential local-orbital code (FPLO) \cite{eschrig2004relativistic}. The generalized gradient approximation (GGA)~\cite{perdew_generalized_1996} for the exchange-correlation functional was used.

{\it VASP.} The projector-augmented wave (PAW) technique was applied~\cite{blochl_projector_1994}. Strong electronic correlations and spin-orbit coupling (SOC) were taken into account via so-called GGA+U+SOC approximation (LDAUTYPE = 1)~\cite{Liechtenstein1995}. The on-site Coulomb repulsion parameter $U$ for Co ions was varied in a range from $5$ to $7$ eV~\cite{Zvereva2016}. The Hund's intra-atomic exchange parameter was taken to be $J_H$ = 0.9 eV. The electronic population numbers for Co were obtained by integration within atomic sphere with radius of $1.302$~\AA. The cutoff energy for plane waves was chosen to be $500$ eV. The integration over the Brillouin zone was performed using the Monkhorst-Pack scheme~\cite{monkhorst_special_1976}. The $7\times 7\times 7$ $k$-mesh was used for conventional DFT calculations (including the Wannier function projection  ~\cite{Mostofi2014} and construction of projected localized orbital - PLO~\cite{Schuler2018}), the $5\times 2\times 3$ $k$-mesh was used for the optimization of the crystal structure, while $3\times 3\times 3$ $k$-grid for calculation of the exchange constants. In two last cases supercells consisting of 4 and 5 formula units (f.u.) were used. 

{\it FPLO.} Hopping integrals were estimated starting from scalar relativistic calculations at the GGA (PBE) level with 12 $\times$ 12 $\times$ 12 $k$-grid. Wannier orbitals were constructed via projection \cite{koepernik2021symmetry}, with the local cubic projection axes chosen such that the $x+y+z$ direction was parallel to the crystallographic $c$-axis, and the angle between each local axis and the corresponding Co-O bond vector was minimized.

{\it Wien2k.} To analyze formation of the bonding and antibonding $e_g$ states due to large hopping between third nearest neighbors, we used the methods described in Ref.~\onlinecite{Foyevtsova2013} based on the eigenstates from the full-potential linearized augmented plane-wave (LAPW) calculations as realized in Wien2k code~\cite{Wien2k,Blaha2020}. We chose the basis-size controlling parameter RK$_{\rm max}$ = 7 and a mesh of 500 {\bf k} points in the first Brillouin zone (FBZ) of the primitive unit cell. The density of states (DOS) was computed with 1000 {\bf k} points in the full Brillouin zone. 

The crystal structure of BaCo$_2$(AsO$_4$)$_2$ was taken from Ref.~\onlinecite{Dordevic2008}. In order to estimate the exchange interaction parameters we consider the anisotropic-exchange model
\begin{eqnarray}
H=\sum_{i \ne j}\vec S_i \mathbb{J}_{ij}  \vec S_j
\end{eqnarray}
where the summation runs twice over each pair of $S=1/2$, and $\mathbb{J}_{ij}$ is a $3\times 3$ matrix. We used the total energy method, described in Ref.~\onlinecite{Xiang2011} in a supercell with 10 Co ions ordered in a honeycomb plane.

\section{Crystal-field splitting and hopping parameters\label{hoppings}} 

All parameters are given in eV.

\begin{itemize}
    \item FPLO results (basis: $d_{xy},d_{yz},d_{xz},d_{z^2},d_{x^2-y^2}$)

On-site Hamiltonian 
\begin{align}
\mathbb{H}_{\rm CFS} = \left(\begin{array}{rrrrr}
 -0.906&  0.044&  0.044& -0.005&  0.034\\
  0.044& -0.906&  0.044& -0.027& -0.021\\
  0.044&  0.044& -0.906&  0.032& -0.013\\
 -0.005& -0.027&  0.032&  0.144&  0.000\\
  0.034& -0.021& -0.013&  0.000&  0.144
 \end{array} \right)
 \nonumber
\end{align}

Nearest neighbor hoppings
\begin{align}
\mathbb{T}_{1} = \left(\begin{array}{rrrrr}
-0.296&  0.044&  0.021&  0.049& -0.003\\
 0.044&  0.062& -0.019&  0.004& -0.019\\
 0.021& -0.019&  0.067&  0.012&  0.038\\
 0.049&  0.004&  0.012& -0.038& -0.004\\
-0.003& -0.019&  0.038& -0.004& -0.041
 \end{array} \right)
 \nonumber
\end{align}

3rd nearest neighbor hoppings
\begin{align}
\mathbb{T}_{3} = \left(\begin{array}{rrrrr}
 -0.038& 0.000& -0.004& 0.021& 0.012\\
 0.000& 0.000& -0.004& -0.001& 0.001\\
 -0.004& -0.004& 0.001& 0.005& 0.009\\
 0.021& -0.001& 0.005& -0.039& -0.001\\
 0.012& 0.001& 0.009& -0.001& 0.124
 \end{array} \right)
 \nonumber
\end{align}

    \item VASP Wannier90 results (basis: $d_{xy},d_{yz},d_{z^2},d_{xz},d_{x^2-y^2}$)
    

On-site Hamiltonian 
\begin{align}
\mathbb{H}_{\rm CFS} = \left(\begin{array}{rrrrr}
 3.953&  0.040& -0.005&  0.040& -0.027\\
 0.040&  3.954&  0.024&  0.040&  0.009\\
-0.005&  0.024&  4.851& -0.022& -0.000\\
 0.040&  0.040& -0.022&  3.954&  0.017\\
-0.027&  0.009& -0.000&  0.017&  4.851
 \end{array} \right)
 \nonumber
\end{align}

Nearest neighbor hoppings
\begin{align}
\mathbb{T}_{1} = \left(\begin{array}{rrrrr}
-0.296&   0.033&   0.049&  0.028&  0.017\\
 0.033&   0.067&  -0.002& -0.020& -0.019\\
 0.049&  -0.002&  -0.037&  0.013&  0.001\\
 0.028&  -0.020&   0.013&  0.067&  0.035\\
-0.017&  -0.019&   0.001&  0.035& -0.042
 \end{array} \right)
 \nonumber
\end{align}

3rd nearest neighbor hoppings
\begin{align}
\mathbb{T}_{3} = \left(\begin{array}{rrrrr}
 -0.039&  0.000&  0.021& -0.004&  0.002\\
  0.000&  0.001& -0.002& -0.003&  0.003\\
  0.021& -0.002& -0.036&  0.007&  0.001\\
 -0.004& -0.003&  0.007&  0.001&  0.013\\
  0.002&  0.003&  0.001&  0.013&  0.125 
 \end{array} \right)
 \nonumber
\end{align}
    
\end{itemize}

We would like to note that on-site energies and hopping parameters obtained using PAW are very close to what one gets using Wannier function projection technique. 

\section{DFT: Bonding-antibonding $t_{2g}$ bands due to a direct hopping}

\begin{figure}[t]
\includegraphics[width=0.45\textwidth]{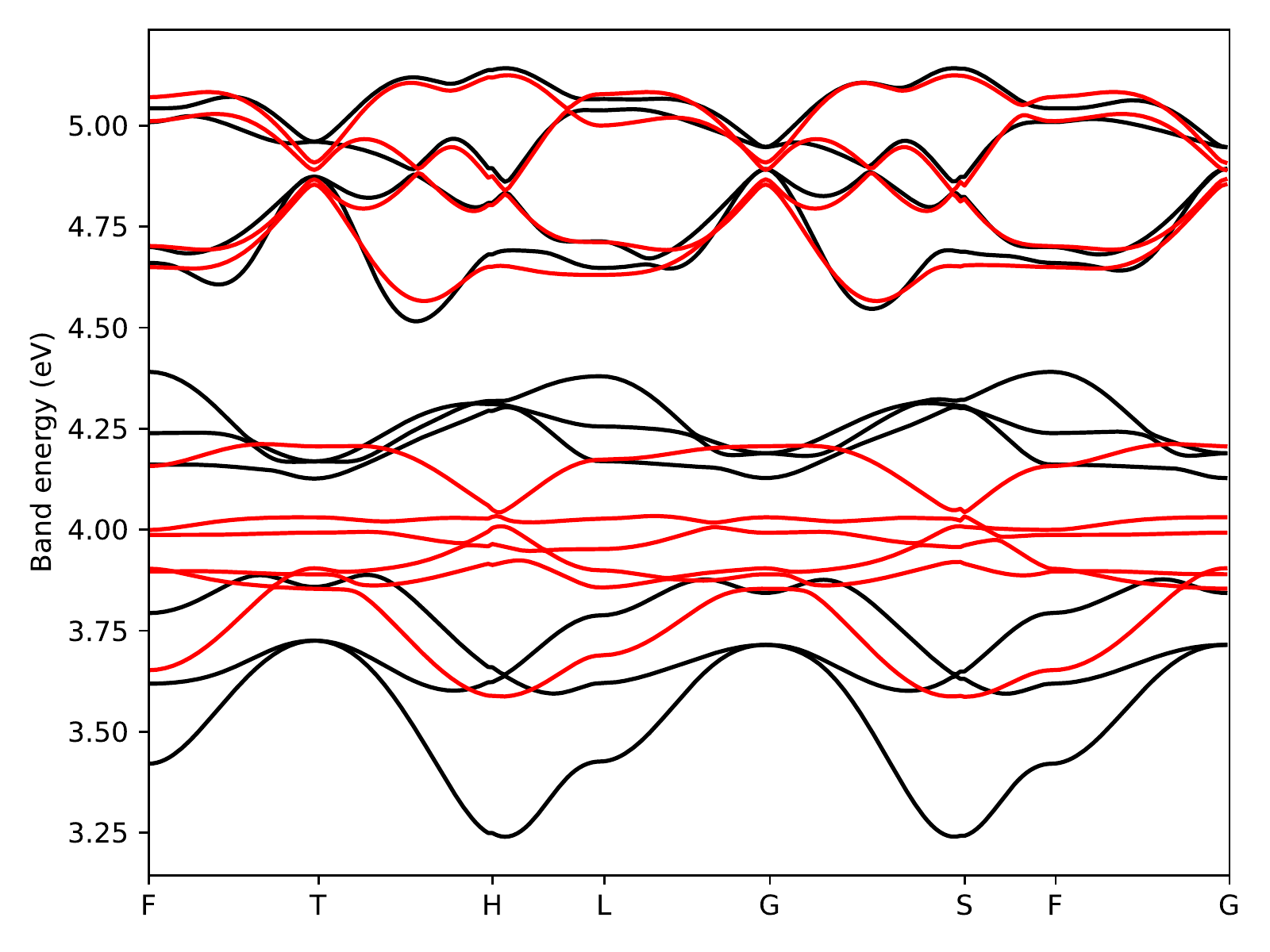}
\caption{\label{t2gbands} Black: band-structure calculated within the tight-binding method using parameters obtained by the Wannier function projection in VASP. Red: the same, but with the $xy/xy$ hopping between nearest neighbors put to zero.}
\end{figure}

In order to clarify importance of the direct hopping between $t_{2g}$ orbitals centered at the neighboring sites we performed Wannier function projection~\cite{Mostofi2014} of DFT Hamiltonian as obtained in VASP. The real space Hamiltonian was transformed to the local coordinate system with axis directed to ligands. Then for each pair nearest Co atoms  $xy/xy$ hopping was put to zero and resulting Hamiltonian rotated back to the global coordinate system and Fourier transformed to the reciprocal space. In Fig.~\ref{t2gbands} resulting band structure (red) is shown together with the one obtained by the tight-binding method using Wannier function projected Hamiltonian (black). One can clearly see that this is the $xy/xy$ hopping, which leads to formation of the gap between two branches of $t_{2g}$ bands.

\section{DFT: Bonding-antibonding $e_{g}$ bands due to third neighbor hopping}
\begin{figure}[ht]
\includegraphics[width=0.45\textwidth]{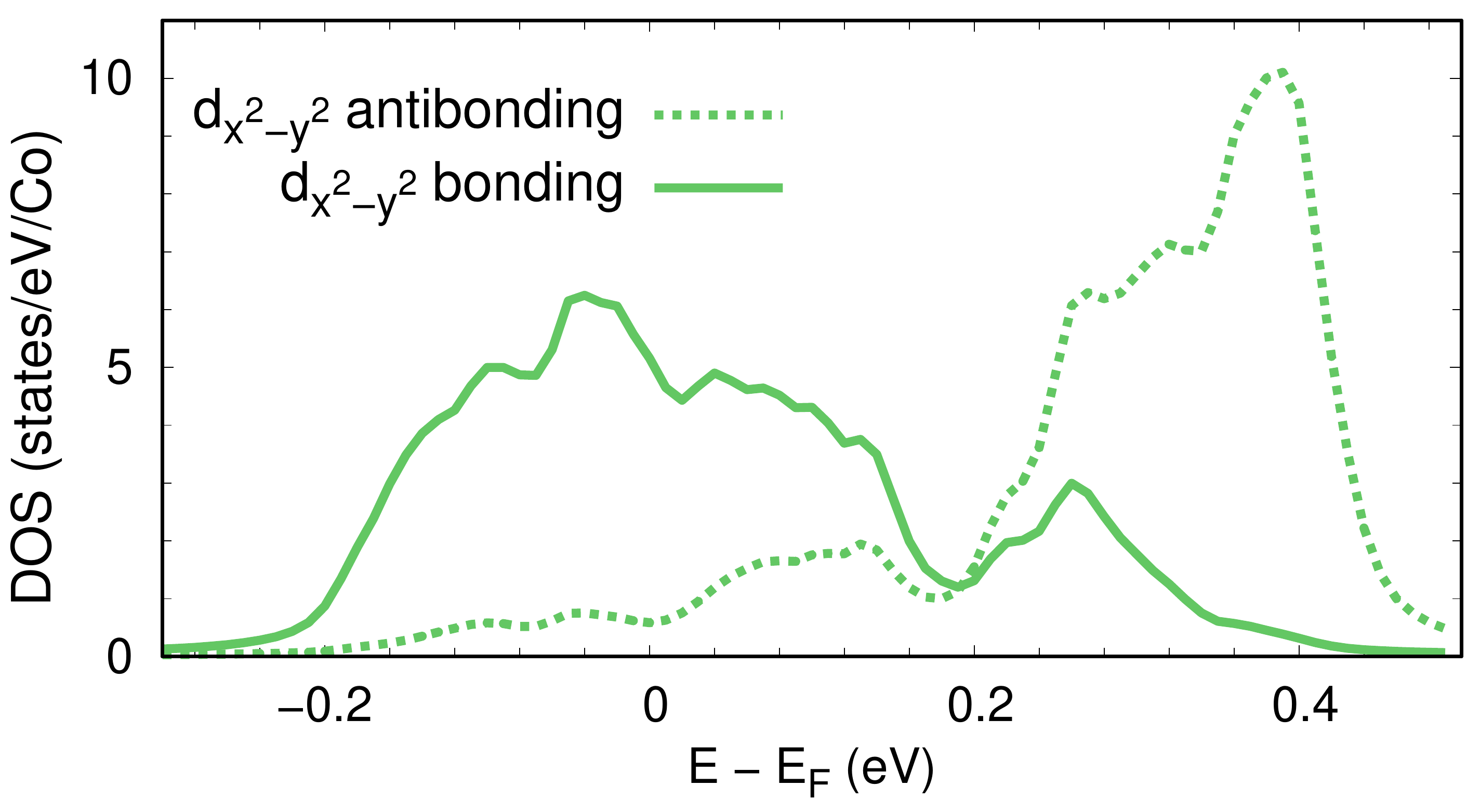}
\caption{\label{qmonn3} Nonrelativistic GGA density of states projected onto bonding and antibonding $d_{x^2-y^2}$ orbitals.}
\end{figure}

There is a substantial hopping between ${x^2-y^2}$ orbitals centered on third nearest neighbors. We therefore project the density of states onto bonding and antibonding $x^2-y^2$ orbitals. This is technically easier to perform in Wien2k code~\cite{Foyevtsova2013} and results of the projection are shown in Fig.~\ref{qmonn3}. 

\section{DFT+U+SOC electronic structure}
\begin{figure}[t!]
\includegraphics[width=0.45\textwidth]{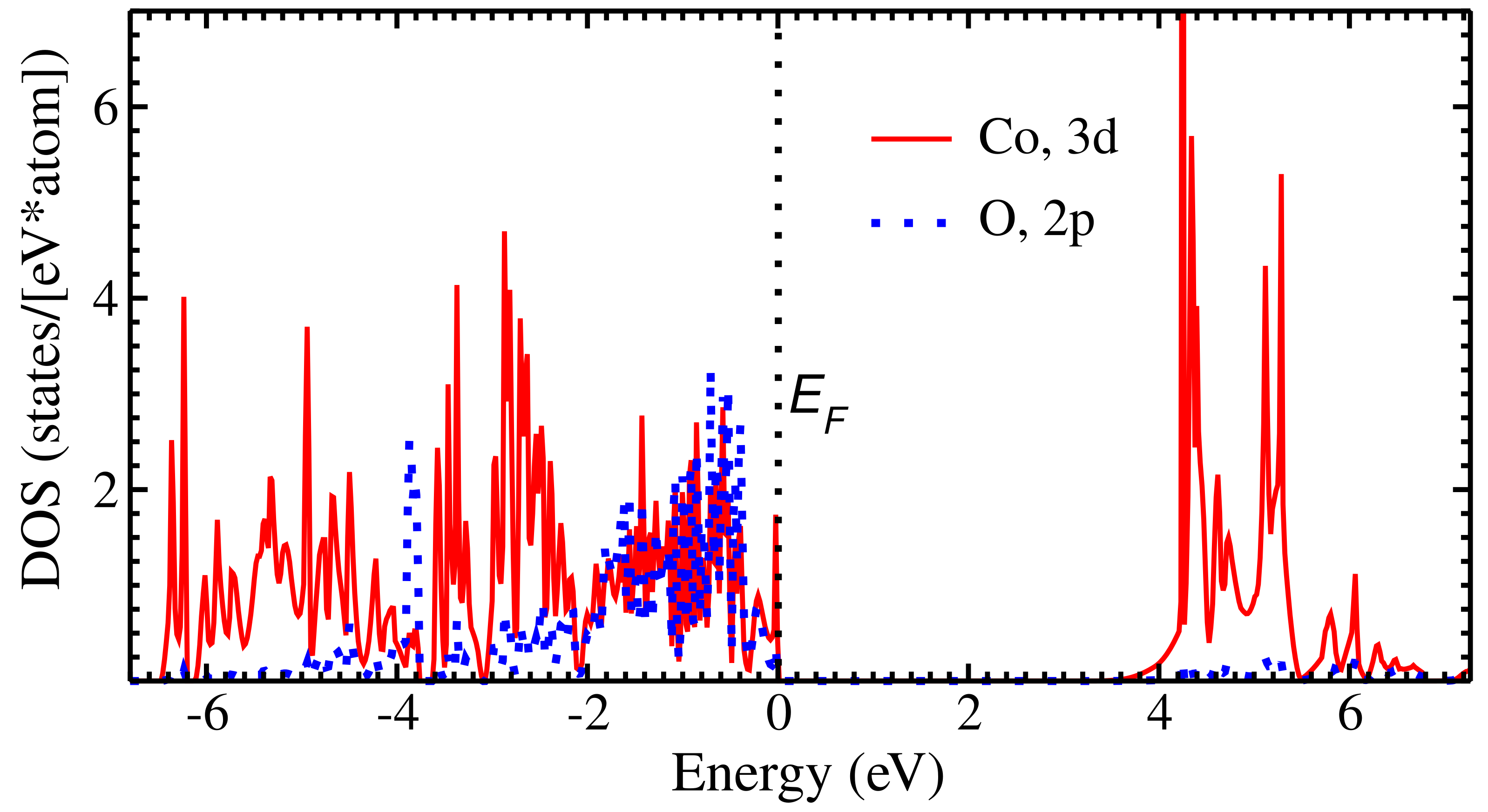}
\caption{\label{DOS} Total density of states as obtained in GGA+U+SOC calculations. Fermi energy is in zero.}
\end{figure}

The electronic structure obtained in GGA+U+SOC calculations is presented in Fig.~\ref{DOS}. 

\section{DFT+U+SOC: Atomic position relaxation}
Relaxation of the atomic positions was performed keeping the volume and unit cell shape constant while variation of the total energy between ionic iterations was more than 10$^{-6}$ eV per unit cell. The hopping parameters obtained by the subsequent Wannier function projection of non-magnetic DFT Hamiltonian are given below.

On-site Hamiltonian 

\begin{align}
\mathbb{H}_{\rm CFS} = \left(\begin{array}{ccccc}
 3.770& 0.038&-0.003& 0.039&-0.013\\
 0.038& 3.770& 0.011& 0.039& 0.002\\
-0.003& 0.011& 5.024&-0.009& 0.000\\
 0.039& 0.039&-0.009& 3.771& 0.008\\
-0.013& 0.002& 0.000& 0.008& 5.022
 \end{array} \right)
 \nonumber
\end{align}

Nearest neighbor hoppings
\begin{align}
\mathbb{T}_{1} = \left(\begin{array}{ccccc}
-0.284& 0.032& 0.058& 0.020&-0.014\\
 0.032& 0.066& 0.001&-0.018&-0.022\\
 0.058& 0.001&-0.048& 0.017& 0.000\\
 0.020&-0.018& 0.017& 0.067& 0.039\\
-0.014&-0.022& 0.000& 0.039&-0.035
 \end{array} \right)
 \nonumber
\end{align}

3rd nearest neighbor hoppings
\begin{align}
\mathbb{T}_{3} = \left(\begin{array}{ccccc}
-0.043& 0.002& 0.026&-0.003& 0.005\\
 0.002& 0.000&-0.004&-0.003& 0.007\\
 0.026&-0.004&-0.039& 0.004& 0.000\\
-0.003&-0.003& 0.004& 0.000& 0.009\\
 0.005& 0.007& 0.000& 0.009& 0.129
 \end{array} \right)
 \nonumber
\end{align}

We did not perform total energy calculations of the exchange parameters in the relaxed structure because of the too large unit cell. Instead the nearest neighbor exchange matrix was calculated using exact diagonalization method. Results are shown in Table \ref{JED}. As can be seen, there are only small modifications compared to the results for the experimental structure presented in the main text, thus highlighting the strong sensitivity of the ground state to the couplings. 


\begin{table}[t!]
\centering
\caption{Exchange interaction parameters (in units of K) computed from exact diagonalization of effective $d$-orbital model. Values in brackets include estimated corrections for omitted ligand exchange processes.}
\begin{tabular}{l | c | c c c}
\hline
\hline
$J_{H,t_{2g}}$&0.7~eV&\multicolumn{3}{c}{0.9~eV}\\
 $U$&3.25 eV & 5 eV  & 6 eV & 7 eV \\
\hline 
 $\text{J}_1$ (K)                & -106 (-126)& -36 (-56)  & -16 (-36)  &  -7.3 (-27) \\
  $K_1$ (K)               & 31 &  12   &5.9 & 2.9 \\ 
$\Gamma_1$ (K)            & 28 (35)&  14 (21) & 8.1 (15) &  4.9 (12)\\ 
  $\Gamma_1^{\prime}$ (K) & 8.8 (16) & 6 (13) & 3.5 (11) & 1.9 (9) \\ 
\hline
\hline
\end{tabular}
\label{JED}
\end{table}


\bibliography{calc-details}